\documentclass[a4paper,11pt]{article}
\pdfoutput=1
\usepackage{jcappub}
\usepackage{slashed}
\usepackage{multirow}
\usepackage[normalem]{ulem}
\graphicspath{{plots/}}
\title{Discovering Heavy Neutral Leptons with the Higgs Boson}
\author{Nicolás Bernal,}
\author{Kuldeep Deka,}
\author{and Marta Losada}

\affiliation{New York University Abu Dhabi\\
PO Box 129188, Saadiyat Island, Abu Dhabi, United Arab Emirates}

\emailAdd{nicolas.bernal@nyu.edu}
\emailAdd{kuldeep.deka@nyu.edu}
\emailAdd{marta.losada@nyu.edu}

\abstract{We study the dominant signatures that arise in Higgs physics at colliders when extending the Standard Model (SM) with a Yukawa interaction to heavy neutral leptons (HNL), while suppressing their mixing to active neutrinos. We focus on the production of HNLs from Higgs bosons that subsequently decay via the Higgs to SM fermions to determine the experimental reach at the LHC detectors and far detectors such as FASER and MATHUSLA. We also determine the impact of precision Higgs constraints on beyond-SM parameters in this scenario.}
\begin{document}
\begin{flushright}
\end{flushright}
\maketitle

%%%%%%%%%%%%%%%%%%%%%%%%%%%%%%%%%%%%%
\section{Introduction}
%%%%%%%%%%%%%%%%%%%%%%%%%%%%%%%%%%%%%
The discovery of the Higgs boson in 2012~\cite{ATLAS:2012yve, CMS:2012qbp} vindicated the Standard Model (SM) to an unprecedented level of accuracy. The myriad of successes of the SM over the years include, but are not limited to, the discovery of gluons~\cite{Barber:1979yr}, the prediction of $W^\pm$/$Z^0$ bosons and their masses~\cite{UA1:1983crd}, and the prediction of charm/top quarks~\cite{E598:1974sol, Roy:1989zd, CDF:1994juo}. These successes are, however, somewhat shadowed due to a few shortcomings which persist till date, among which neutrino mass, dark matter, matter-antimatter asymmetry, and a few anomalies in the experimental observables are the most vexing ones. There is a plethora of beyond-SM (BSM) scenarios that have been proposed to address these shortcomings. The way forward now is to meaningfully constrain these scenarios through precision Higgs-physics observables, be it at the HL-LHC or future Higgs factories.

Following the confirmation of neutrino masses through neutrino oscillation experiments, a natural question arises regarding whether they are somehow linked to the Higgs mechanism, which imparts mass to all other elementary particles through the electroweak symmetry breaking. One plausible extension of the SM to address this inquiry involves the introduction of heavy neutral leptons (HNLs). These sterile neutrinos, being singlets under the SM gauge group, interact with the Higgs boson and active SM neutrinos through Yukawa interactions. Moreover, {\it depending on the model,} they can undergo mixing with SM neutrinos following the electroweak phase transition, once the Higgs acquires a vacuum expectation value. For detailed reviews, see Refs.~\cite{Abazajian:2012ys, Abdullahi:2022jlv}.
Non-zero neutrino masses can be realized through various types of mechanisms, popular among which include seesaw, scotogenic, etc.~\cite{Minkowski:1977sc, Gell-Mann:1979vob, Yanagida:1979as, Mohapatra:1979ia, Glashow:1979nm, Schechter:1980gr, Schechter:1981cv, Foot:1988aq, Ma:1998dn, Choudhury:2020cpm}. Apart from the generation of neutrino masses, these HNLs can also help in the dynamical generation of the baryon asymmetry of the universe via leptogenesis~\cite{Fukugita:1986hr, Akhmedov:1998qx, Asaka:2005pn, Davidson:2008bu, Hambye:2016sby, Deka:2021koh} and may even be a viable candidate for explaining the observed dark-matter relic abundance~\cite{Dodelson:1993je, Shi:1998km, Abazajian:2001nj}. Interestingly, all three of these problems can be solved within the framework of neutrino minimal SM by the introduction of three HNLs~\cite{Asaka:2005an, Asaka:2005pn}.

However, instead of focusing on particular models, here we follow a model-independent approach and study collider experimental tests of HNLs.
Many experimental searches have focused on the high-mass regime, where HNLs are produced directly or in some prompt decay channels~\cite{Abreu:1996pa, Aad:2011vj, Chatrchyan:2012fla, Khachatryan:2015gha, Aad:2015xaa, Khachatryan:2016olu, Sirunyan:2018mtv}, with numerous dedicated analyzes, for example, in Refs.~\cite{delAguila:2007qnc, delAguila:2008cj, delAguila:2008hw, Atre:2009rg, BhupalDev:2012zg, Dev:2013wba, Das:2014jxa, Alva:2014gxa, Deppisch:2015qwa, Banerjee:2015gca, Arganda:2015ija, Das:2015toa, Degrande:2016aje, Mitra:2016kov, Das:2017zjc, Das:2017rsu, Ruiz:2017yyf, Cai:2017mow, Accomando:2017qcs, Drewes:2018gkc, Pascoli:2018rsg, Bhaskar:2023xkm}.

Another possibility occurs in the parameter space where HNLs can decay with a sizable displacement in LHC detectors, featuring a displaced vertex that is a distinctive signature of their existence~\cite{Gronau:1984ct}. Recently, several studies have promoted the use of these types of dedicated LHC searches for HNL with associated charged leptons~\cite{Nemevsek:2011hz, Helo:2013esa, Izaguirre:2015pga, Dube:2017jgo, Cottin:2018kmq, Cottin:2018nms, Dib:2018iyr, Nemevsek:2018bbt, Abada:2018sfh, Marcano:2018fto, Abada:2018ulr}, from Higgs decays~\cite{Maiezza:2015lza, Gago:2015vma, Accomando:2016rpc, Nemevsek:2016enw, Caputo:2017pit, Deppisch:2018eth, Liu:2018wte}, or for LHCb~\cite{Antusch:2017hhu}. There is also a potential to search for such signatures at DUNE~\cite{Adams:2013qkq, Gunther:2023vmz}, IceCube~\cite{Coloma:2017ppo}, future lepton colliders~\cite{Blondel:2014bra, Antusch:2016vyf} and SHiP~\cite{Alekhin:2015byh, Anelli:2015pba}, the latter expected to greatly improve the sensitivity to HNL below the $c$-quark mass, with HNLs abundantly produced by meson decays~\cite{Bonivento:2013jag}. CMS and ATLAS have already performed some of such analyzes through various channels~\cite{CMS:2015qur, CMS:2018iaf, ATLAS:2019kpx, CMS:2022fut, ATLAS:2022atq}.

In the same spirit, several specific far detectors have been proposed at the LHC, with the aim of detecting long-lived particles, including FASER~\cite{Feng:2017uoz, FASER:2018eoc}, MoEDAL-MAPP~\cite{Pinfold:2019nqj, Pinfold:2019zwp}, MATHUSLA~\cite{Chou:2016lxi, Curtin:2018mvb, MATHUSLA:2020uve}, ANUBIS~\cite{Bauer:2019vqk}, CODEX-b~\cite{Gligorov:2017nwh}, and FACET~\cite{Cerci:2021nlb}. Among them, FASER and MoEDAL-MAPP are already in operation, and others are still in discussion. The particular scenario where HNLs are long-lived particles has been extensively discussed in the recent literature~\cite{Caputo:2016ojx, Jana:2018rdf, Kling:2018wct, Helo:2018qej, Dercks:2018wum, Deppisch:2019kvs, Frank:2019pgk, Hirsch:2020klk, Li:2023dbs, Deppisch:2023sga}.

It is important to note that all analyses conducted so far have focused on the search for HNLs through their mixing with SM active neutrinos. However, the presence of significant mixing and the production/decay of HNLs via this mixing highly depends on the model and are not always guaranteed. As discussed in the appendix, there could be scenarios where the Yukawa interactions dominate and contribute solely to the HNL phenomenology. Therefore, as an alternative to the conventional approach, we propose a novel and complementary method that exploits their Yukawa coupling to the SM Higgs boson. We adopt a cautious stance and treat Yukawa couplings and mixing angles as independent parameters to avoid assumptions. For simplicity and as an initial step, we assume that all mixing angles are very small, focusing on the role of trilinear Yukawa couplings. This approach could also provide information on the Higgs mechanism and the electroweak symmetry breaking.

The paper is structured as follows. In Section~\ref{sec:framework}, the model and the setup of the analysis are presented. Then, in Section~\ref{sec:HNL}, the production and decay modes of HNLs through Higgs interactions are described, together with the constraints from precision Higgs measurements. Sections~\ref{sec:Prompt}, \ref{sec:displaced}, and~\ref{sec:LLP} are devoted to the analysis of searches for HNLs that decay promptly, produce a displaced vertex, or decay in far detectors, respectively. Finally, we conclude in Section~\ref{sec:concl}.

%%%%%%%%%%%%%%%%%%%%%%%%%%%%%%%%%%%%%
\section{Set-up} \label{sec:framework}
%%%%%%%%%%%%%%%%%%%%%%%%%%%%%%%%%%%%%
In this work, we extend the SM by introducing $n$ HNLs denoted as $\widetilde{N}$, which could be Dirac or Majorana particles. The relevant Yukawa interaction term is expressed as $y_{N \alpha}\, \overline{\widetilde{N}}\, \widetilde{H}^\dagger\, L_\alpha$, where $H$ represents the SM Higgs doublet, $\widetilde{H} \equiv i \sigma^2 H^*$ (with $\sigma^2$ being the second Pauli spin matrix), $L_\alpha$ are the SM lepton doublets for $\alpha = e$, $\mu$, $\tau$ and $y_{N \alpha}$ are the Yukawa couplings between the HNLs and the SM leptons. We denote the mass of the HNL as $m_{N}$, without explicitly writing the mass term, as it differs slightly depending on whether $\widetilde{N}$ is a Dirac or Majorana particle.

Following electroweak symmetry breaking, the Higgs field acquires a vacuum expectation value (vev) $v_h \simeq 246$ GeV, leading to Yukawa interactions generating Dirac mass-like terms $(M_D)_{N\alpha} \equiv {y_{N\alpha}v_h}/{\sqrt{2}}$. Similar terms can also arise if additional scalar fields are present in the model, which also acquire vevs. The physical mass eigenstates for the three SM neutrinos $\nu$ and the $n$ HNLs are determined by diagonalizing the full mass matrix specific to the model. To fit neutrino oscillation data, at least $n \geq 2$ HNLs are required~\cite{Esteban:2020cvm, deSalas:2020pgw}. Looking to accommodate sub-eV neutrino masses~\cite{KATRIN:2021uub} imposes constraints on the parameter space of the model. The diagonalization to the mass basis could also cause mixings between HNLs and SM neutrinos. If present, the mixings result in HNLs inheriting interactions with other SM particles, especially the $W^\pm$ and $Z$ bosons. For a Type-I seesaw mechanism~\cite{Minkowski:1977sc, Gell-Mann:1979vob, Yanagida:1979as, Mohapatra:1979ia, Glashow:1979nm, Schechter:1980gr, Schechter:1981cv}, the mixing scales as $y_{N \alpha} v_h/m_N$. However, in inverse seesaw and double-seesaw models~\cite{Dias:2012xp,Fraser:2014yha}, the mixing contributions can be smaller due to the presence of an additional mass scale. There are also cases where the mixing contributions can be tuned to be precisely zero~\cite{Ma:2014qra, CentellesChulia:2016rms}, as explicitly discussed in the Appendix~\ref{appendix}. Overall, it is important to emphasize that, while there are well-motivated models linking Yukawa couplings and mixing angles, these relations should not be assumed to hold universally.

For an agnostic perspective, free from specific model assumptions, we consider Yukawa couplings and mixing angles as independent parameters. As a first step, we focus on a single HNL and its production and decay through interactions with the Higgs boson, assuming negligible mixing with the SM neutrinos, resulting in minimal coupling of HNLs to $W$ and $Z$ bosons.\footnote{For example, with zero mixings at tree-level, $W \rightarrow N \ell$ can still arise at one-loop through a virtual Higgs, but is suppressed by the loop factor and the square of the Yukawa of the lepton.}

%%%%%%%%%%%%%%%%%%%%%%%%%%%%%%%%%%%%%
\section{Production and Decays of HNLs} \label{sec:HNL}
%%%%%%%%%%%%%%%%%%%%%%%%%%%%%%%%%%%%%
In this work, we consider the production of HNLs through the decay of the Higgs boson, which implies that $m_N$ has to be in the GeV ballpark. Much heavier particles could not be kinematically produced by Higgs decays; lighter HNLs, below a few GeVs, are mainly produced by meson decays~\cite{Bondarenko:2018ptm}.

For the following phenomenological analysis, we have used a modified version of the \texttt{FeynRules}~\cite{Degrande:2011ua, Alloul:2013bka} model \texttt{HeavyN}~\cite{Atre:2009rg, Alva:2014gxa, Degrande:2016aje}, with a single HNL and only Yukawa couplings. Signal and background events are generated using \texttt{MADGRAPH5\_aMC@NLO} v2.9.16~\cite{Alwall:2014hca, Frederix:2018nkq} interfaced with \texttt{PYTHIA} 8~\cite{Bierlich:2022pfr} for parton showering and fragmentation. The events are then passed through \texttt{Delphes} 3.5~\cite{deFavereau:2013fsa} in order to implement detector effects and various reconstruction algorithms.

%%%%%%%%%%%%%%%%%%%%%%%%%%%%%%%%%%%%%
\subsection{Production of HNLs}
%%%%%%%%%%%%%%%%%%%%%%%%%%%%%%%%%%%%%
Here we consider the production of HNLs through Higgs decays.
At the LHC, Higgs bosons are produced mainly by the gluon-gluon fusion mechanism (ggF) mediated by triangular loops of heavy quarks, with a cross section $\sigma_\text{ggF} \simeq 54.8$~pb at $\sqrt{s} = 14$~TeV~\cite{ggF}.
The second most important channel corresponds to vector-boson fusion (VBF), with a cross section $\sigma_\text{VBF} \simeq 4.26$~pb also at $\sqrt{s} = 14$~TeV~\cite{VBF}.
Other channels such as the associated production with massive gauge bosons or heavy-quark pairs are further suppressed~\cite{Djouadi:2005gi}.

\begin{figure}[t!]
    \def\sepf{0.49}
    \centering
    \includegraphics[width=\sepf\columnwidth]{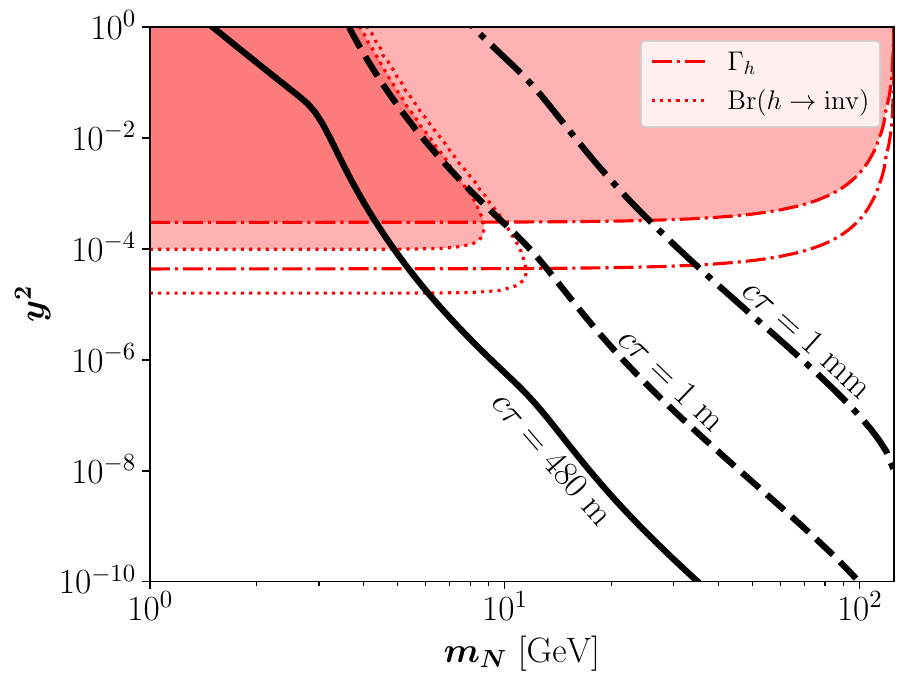}
    \includegraphics[width=\sepf\columnwidth]{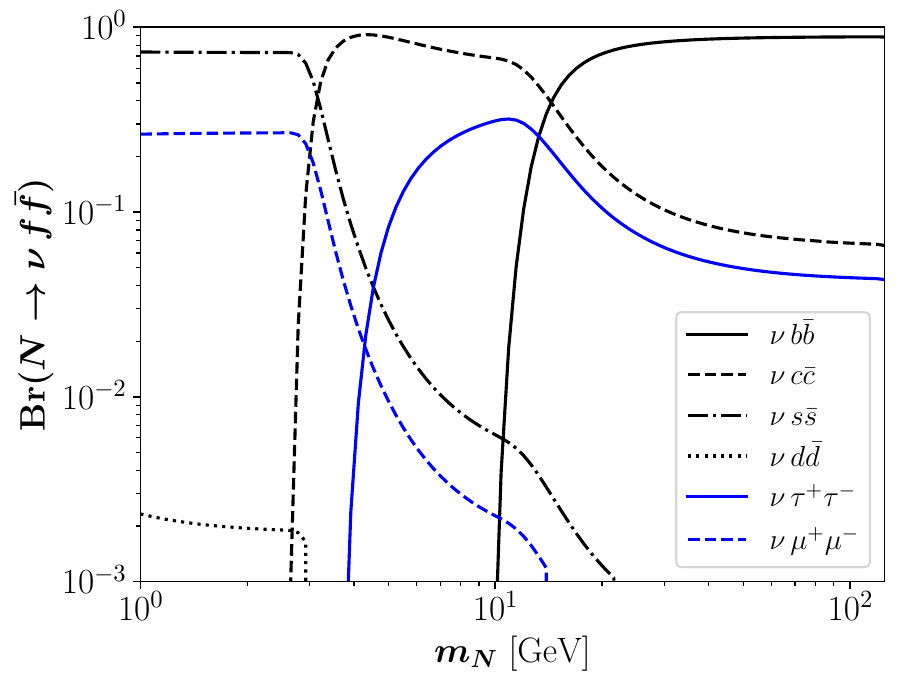}
    \caption{Left: Contours for the HNL decay lengths $c\tau = 1$~mm, 1~m, and 480~m. The red areas are in tension with the invisible decay of the Higgs or its total decay width, whereas the red dotted and dashed dotted correspond to projected sensitivities at HL-LHC. Right: Branching ratios for the 3-body decay of the HNL, summing over all neutrino flavors.}
    \label{fig:Br}
\end{figure}

The partial decay width of the Higgs boson into a HNL and an active neutrino is given by
\begin{equation} \label{eq:Higgsdec}
    \Gamma(h\to N \nu) = \frac{y^2}{8 \pi}\, m_h \left[1 - \left(\frac{m_N}{m_h}\right)^2\right]^2,
\end{equation}
where $m_h \simeq 125$~GeV is the Higgs mass.
It is interesting to note that the neutrino flavor $\nu_\alpha$ depends on the Yukawa $y_{N\alpha}$. However, since the LHC is insensitive to this flavor, the coupling $y$ appearing in Eq.~\eqref{eq:Higgsdec} must be understood as the sum of all Yukawa couplings, that is, $y^2 \equiv y_{Ne}^2 + y_{N\mu}^2 + y_{N\tau}^2$.
This new decay mode contributes to the total decay width of the Higgs, which in the SM is $\Gamma_h \simeq $~4.1~MeV~\cite{LHCHiggsCrossSectionWorkingGroup:2016ypw}, and has been measured at the LHC to be $\Gamma_h = 3.2^{+2.4}_{-1.7}$~MeV~\cite{ParticleDataGroup:2022pth}.
HL-LHC is expected to improve this measurement to a precision of $5.3\%$~\cite{deBlas:2022aow}. In the left panel of Fig.~\ref{fig:Br}, the red dashed dotted region labeled $\Gamma_h$ shows the parameter space in tension with measurements of the Higgs total decay width and projections for HL-LHC.
%%%%%%%%%%%%%%%%%%%%%%%%%%%%%%%%%%%%%%%%%%%%%%%%%%%
%%%%%%%%%%%%%%%%%%%%%%%%%%%%%%%%%%%%%%%%%% 

%%%%%%%%%%%%%%%%%%%%%%%%%%%%%%%%%%%%%
\subsection{Decays of HNLs}
%%%%%%%%%%%%%%%%%%%%%%%%%%%%%%%%%%%%%
In the present scenario, HNLs can only decay to an active neutrino and an off-shell Higgs that later decays to a  fermion-antifermion pair. Away from kinematical thresholds, each individual partial decay width scales as
\begin{equation} \label{eq:Ndec}
    \Gamma(N \to \nu f \bar f) \propto y^2\, \frac{m_f^2\, m_N^5}{m_h^6}\,,
\end{equation}
where $m_f$ corresponds to the mass of the fermion $f$.
Furthermore, due to the mass hierarchy, the total decay width is dominated by the contribution of $s$ quarks (1~GeV $< m_N \lesssim 3$~GeV), $c$ quarks (3~GeV $\lesssim m_N \lesssim 15$~GeV) or $b$ quarks (15~GeV $\lesssim m_N < 125$~GeV), with corresponding branching fractions Br$(N \to \nu f \bar f)$  shown in the right panel of Fig.~\ref{fig:Br}. It is interesting to note that the branching fractions depend only on $m_N$ (and not on $y$), as the flavor of the active neutrino is not measurable at the LHC.
Furthermore, the lifetime of HNLs can span a wide range and therefore can have prompt decays ($c\tau \lesssim 1$~mm), displaced decays ($\mathcal{O}(1)$~mm $\lesssim c\tau \lesssim \mathcal{O}(1)$~m), or even decays outside of the ATLAS or CMS detectors ($\mathcal{O}(1)$~m $\lesssim c\tau$), where $c\tau$ corresponds to the decay length. Additionally, if $c\tau \sim \mathcal{O}(100)$~m, HNLs can decay in far detectors such as FASER or MATHUSLA. The left panel of Fig.~\ref{fig:Br} shows contours with thick black lines for $c\tau$ in the plane $[m_N,\, y^2]$.

It is interesting to note that in our framework the Higgs boson does not have any additional invisible decay mode.\footnote{In the SM, the main invisible decay mode of the Higgs corresponds to $h \to Z Z^* \to 4\nu$, with a branching fraction $\sim 0.1\%$~\cite{ParticleDataGroup:2022pth}. The decay $h \to N \nu$ with $N \to \nu \nu \nu$ can only occur through the mediation of a $Z$ boson, and therefore it is not present in our scenario where the mixing angles are zero.}
However, one can still have an effective contribution to its invisible decay, if the HNL decays outside the detector.
The probability of a HNL to decay outside the detector is given by
\begin{equation}
    \mathcal{P} = \frac{1}{\mathcal{N}_\text{ev}}\, \sum_{i=1}^{\mathcal{N}_\text{ev}} \Theta\left(d_{xy}^{(i)}-L_{xy}\right) \Theta\left(d_z^{(i)}-L_z\right)
\end{equation}
with $\Theta$ being the Heaviside step function,
\begin{align}
    d_{xy}^{(i)} &= c \tau\, \frac{\sqrt{\left(p_x^{(i)}\right)^2 + \left(p_y^{(i)}\right)^2}}{m_N}\,,\\
    d_z^{(i)} &= c \tau\, \frac{|p_z^{(i)}|}{m_N}\,,
\end{align}
are the transversal and the longitudinal decay lengths in the laboratory frame, and $(p_x,\, p_y,\, p_z)$ the 3-momentum of the HNL, for a large number of events $\mathcal{N}_\text{ev}$. Therefore, the corresponding Higgs boson invisible decay branching ratio $\text{Br}(h \to \text{inv})$ is given by
\begin{equation}
    \text{Br}(h \to \text{inv}) = \frac{\mathcal{P}\, \Gamma(h \to N\, \nu)}{\Gamma(h \to N\, \nu) + \Gamma_h}\,,
\end{equation}
and has being measured by the ATLAS collaboration to be $\text{Br}(h \to \text{inv}) < 10.7\%$~\cite{ATLAS:2023tkt} and the CMS collaboration $\text{Br}(h \to \text{inv}) < 15\%$~\cite{CMS:2023sdw} at 95\% CL. Furthermore, future measurements at the HL-LHC are expected to strengthen this bound to $\text{Br}(h \to \text{inv}) \lesssim 1.9\%$~\cite{EuropeanStrategyforParticlePhysicsPreparatoryGroup:2019qin}.
The physical dimensions of the ATLAS detector are about $L_{xy} \simeq 11$~m in radius and $2\, L_z \simeq 44$~m in length, up to the muon spectrometer~\cite{ATLAS:1997ad}.
In the left panel of Fig.~\ref{fig:Br}, the red dotted region labeled $\text{Br}(h \to \text{inv})$ shows the parameter space in tension with the constraint from the invisible decay of the Higgs and projections for HL-LHC.

%%%%%%%%%%%%%%%%%%%%%%%%%%%%%%%%%%%%%%%%%%%%%%%%%%%%%%%%%%
\section{Prompt Decays} \label{sec:Prompt}
%%%%%%%%%%%%%%%%%%%%%%%%%%%%%%%%%%%%%%%%%%%%%%%%%%%%%%%%%%
In this section, we first look at single Higgs bosons produced through VBF in the quest for reducing the background by exploiting the characteristic features of its two hard forward jets. We will use an integrated luminosity $\mathcal{L} = 3$~ab$^{-1}$ and a center-of-mass energy $\sqrt{s} = 14$~TeV.

VBF has a very peculiar topology, with two leading jets ($j_1$ and $j_2$) typically present in the forward region and reside in opposite hemispheres of the detector. This results in a large pseudo-rapidity separation $\Delta \eta_{j_1 j_2}$ and a large invariant mass $m_{j_1 j_2}$. The Higgs boson resulting from VBF also tends to have a significant transverse momentum $p_T$, which requires that the azimuthal separation between the two leading jets $\Delta \phi_{j_1 j_2}$ be small. This is in contrast to what is expected for QCD multijet events and hence can be used as a strategy to suppress such large backgrounds. The Higgs particle then decays to a HNL and a SM neutrino, given by Eq.~\eqref{eq:Higgsdec}. The HNL can then further decay to a pair of $b\bar{b}$ and an active SM neutrino mediated by an off-shell Higgs. As shown in the right panel of Fig.~\ref{fig:Br}, above the $b\bar{b}$ mass threshold, this channel has the highest branching ratio. The main background that contributes to this process is the SM VBF Higgs production itself, where the Higgs decays subsequently to a $b\bar{b}$ pair. The ggF process for Higgs production, with the Higgs decaying to $b\bar{b}$, is also an important background to consider. Although the ggF event topology is very different from that of VBF, there is a non-vanishing probability of the two initial gluons radiating off two hard forward jets that mimic the VBF topology. The effect is enhanced by the fact that the production cross section of ggF is one order of magnitude higher than that of VBF. This non-negligible contribution also motivated in our study the inclusion of the ggF production channel into the signal events of the process we are interested in.

Among other backgrounds, the $t \bar{t}$ + jets case is important because of the automatic presence of two $b$ jets along with other light jets that could mimic the two forward jets in VBF. The cross section is also quite large (990~pb at NNLO~\cite{Muselli:2015kba}) compared to the VBF and ggF processes. Additionally, $b\bar{b}$ + jets with its huge cross section ($\sim 10^5$~pb~\cite{ATLAS:2011ac}) can also be relevant if the associated jets are in the forward region. 

Initial- and final-state radiation in VBF processes can also result in additional jets. Specific requirements on such jets also help to remove QCD multi-jet backgrounds, including $t \bar{t}$ + jets. One such metric used to examine whether a subleading jet can be associated with having been emitted from either of the two primary leading jets in the context of a VBF process is the comparison between the invariant mass of the subleading jet and one of the two primary leading jets. This comparison is made with respect to the invariant mass $m_{j_1 j_2}$. Specifically, it checks whether the invariant mass of the subleading jet $j_i$ (with $i \geq 3$) and that of one of the two leading jets is smaller compared to $m_{j_1 j_2}$, and is given by
\begin{equation}
    m^\text{rel}_{j_i} \equiv \frac{\text{min}(m_{j_1 j_i},\, m_{j_2 j_i})}{m_{j_1 j_2}}\,.
\end{equation}
Small values of $m^\text{rel}_{j_i} $ indicate that the additional jet is compatible with the final-state radiation.

%%%%%%%%%%%%%%%%%%%%%%%%%%%%%%%%%%%%%%%%%%%%%%%%%%%%%%%%%%%%%%%%%%%%%%%
\begin{figure}[t!]
    \def\sepf{0.49}
    \centering
    \includegraphics[width=\sepf\columnwidth]{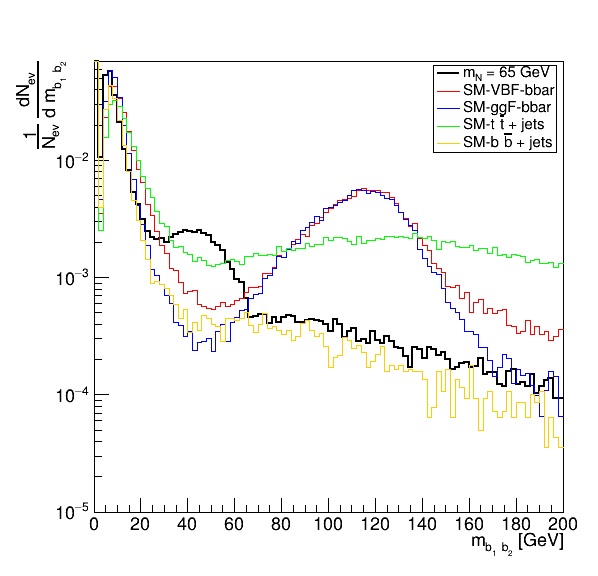}
    \includegraphics[width=\sepf\columnwidth]{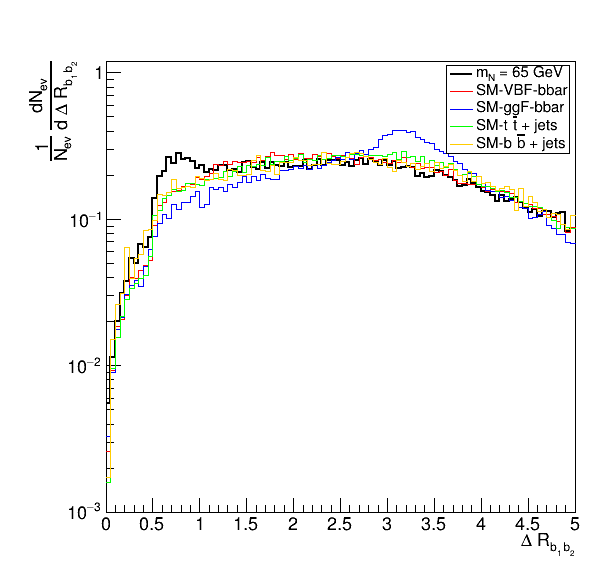}
    \includegraphics[width=\sepf\columnwidth]{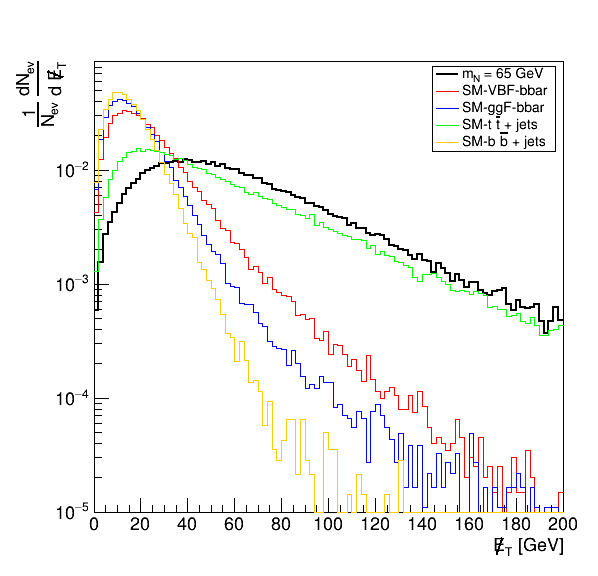}
    \caption{Kinematic distributions for the invariant mass $m_{b_1 b_2}$ and the separation $\Delta R_{b_1 b_2}$ of the two $b$-tagged jets, and the missing transverse energy $\slashed{E}_T$, for $m_N = 65$~GeV.}
    \label{fig:kinematics1}
\end{figure} 
%%%%%%%%%%%%%%%%%%%%%%%%%%%%%%%%%%%%%%%%%%%%%%%%%%%%%%%%%%%%%%%%%%%%%%%
Finally, the two $b$ quarks arising from the HNL decay are expected to carry an invariant mass closer to the actual HNL mass. Although the entire HNL mass cannot be resolved due to the presence of two different sources of missing transverse energy $\slashed{E}_T$ (i.e. the two active neutrinos in the final state), a cut on the invariant mass of the two $b$ quarks around the HNL mass still gives some handle to reduce the SM VBF, ggF, and $t \bar{t}$ + jets backgrounds. A cut on the maximum $\Delta R$ of the two $b$ jets helps to further reduce the $t \bar{t}$ + jets and SM-ggF background events, as their associated $b$ jets tend to have a larger separation.\footnote{$\Delta R$ denotes the distance in the pseudorapidity-azimuth plane, namely $\Delta R \equiv \sqrt{(\Delta \eta)^2 + (\Delta \phi)^2}$.} The two upper plots of Fig.~\ref{fig:kinematics1} show the distributions for the invariant mass $m_{b_1 b_2}$ and the separation $\Delta R_{b_1 b_2}$ of the two $b$-tagged jets for the signal corresponding to $m_N = 65$~GeV (thick black lines) and the four SM background processes discussed (thin colored lines). As already discussed, the two neutrinos in the final state result in a substantial amount of missing energy, as shown in the lower plot of Fig.~\ref{fig:kinematics1}.

Taking cues from the behavior of the signal and backgrounds, we employ the following selection criteria:
\begin{itemize}
    \item{S1:} We demand that the event contains no charged lepton candidates nor photons.
    % , namely $n_\text{lep} = 0$ and $n_\gamma = 0$.
    \item{S2:} We require at least two jets that are not $b$ tagged and at least two $b$-tagged jets.
    % : $n_b \geq 2$, $n_j \geq 2$.
    \item{S3:} We want the two leading non-$b$ jets to satisfy: $p_{T_{j_1}} > 60$~GeV and $p_{T_{j_2}} > 40$~GeV. We also want the total scalar sum of all non-$b$ jets to be $H_T > 140$~GeV.
    \item{S4:} Among the observables derived from the two leading non-$b$ jets, we demand that they follow the VBF topology by requiring them to be in opposite longitudinal hemispheres ($\eta_{j_1} \times \eta_{j_2} < 0$), to have a large pseudorapidity separation ($\Delta \eta_{j_1 j_2} > 3.5$) and a large invariant mass ($m_{j_1 j_2} > 500$~GeV). In addition, they should not be back-to-back ($\Delta \phi_{j_1 j_2} < 2.5$).
    \item{S5:} Any existing third or fourth non-$b$ jet must have $m^\text{rel}_{j_i} < 0.08$.
    \item{S6:} A cut on the missing transverse energy is helpful in reducing the background arising from the decays of SM Higgs to two $b$ jets. We demand $\slashed{E}_T > 50$~GeV. 
    \item{S7:} Finally, we demand a cut on the invariant mass of the two leading $b$-tagged jets given by $0.2\, m_N \leq m_{b\bar{b}} \leq 1.6\, m_N$. We also put a cut on the separation between the two $b$-tagged jets given by $\Delta R_{b \bar{b}} \leq 2.5$. 
\end{itemize}
An example of the cut flow after the implementation of the above cuts for $m_N =65$~GeV is shown in Table~\ref{tab:cutflow}.
%%%%%%%%%%%%%%%%%%%%%%%%%%%%%%%%%%%%%%%%%%%%%%%%%%%%%%%%%%%%%%%%%%%%%%%
\renewcommand{\arraystretch}{1.2}
\begin{table}[t!]
    \begin{small}
        \begin{center}
            \begin{tabular}{|c||c|c||c|c|c|c|}
                \hline
                \multirow{2}*{\bf Cut flow} & \multicolumn{2}{|c||}{\bf Signal} & \multicolumn{4}{|c|}{{\bf Backgrounds}} \\ \cline{2-7}
                 & {\bf VBF} & {\bf ggF} & {\bf SM-VBF} & {\bf SM-ggF} & {\bf $\boldsymbol{t\bar{t}}$ + jets} & {\bf $\boldsymbol{b\bar{b}}$ + jets}\\ \hline\hline
                $\boldsymbol{S_1}$ and $\boldsymbol{S_2}$ & $5.96 \times 10^{-2} $ & $1.24 \times 10^{-2} $& 0.22 & $6.1 \times 10^{-2} $ & 0.3 & $1.66 \times 10^{-2}$ \\ \hline
                $\boldsymbol{S_3}$ &$3.43 \times 10^{-2}$ &$5.92 \times 10^{-3} $ & $0.12$  & $2.2 \times 10^{-2}$ & 0.24 &$4.29 \times 10^{-3}$ \\ \hline
                $\boldsymbol{S_4}$ &$2.72 \times 10^{-3}$ &$1.1 \times 10^{-4} $ & $8.16 \times 10^{-3} $  & $5.1 \times 10^{-4} $ & $2.41 \times 10^{-3} $ & $8 \times 10^{-5} $ \\ \hline 
                $\boldsymbol{S_5}$ &$1.56 \times 10^{-3}$ &$ 8.0 \times 10^{-5} $& $4.48 \times 10^{-3} $&$3.4 \times 10^{-4} $& $1.3 \times 10^{-4} $& $10^{-5}$\\ \hline
                $\boldsymbol{S_6}$ &$1.2 \times 10^{-3}$ &$7 \times 10^{-5} $ &$5.4 \times 10^{-4} $ &$3 \times 10^{-5}$ &$7 \times 10^{-5} $ & 0 \\ \hline
                $\boldsymbol{S_7}$ & $8.8 \times 10^{-4}$ &$3 \times 10^{-5} $ &$2.2 \times 10^{-4} $ &$2 \times 10^{-5}$&0 &  0\\ \hline 
            \end{tabular}
            \caption{Cut flow of fraction of surviving events for $m_N = 65$~GeV.
            }
            \label{tab:cutflow}
        \end{center}
    \end{small}
\end{table}
\renewcommand{\arraystretch}{1}
%%%%%%%%%%%%%%%%%%%%%%%%%%%%%%%%%%%%%%%%%%%%%%%%%%%%%%%%%%%%%%%%%%%%%%%

\begin{figure}[t!]
    \def\sepf{0.49}
    \centering
    \includegraphics[width=\sepf\columnwidth]{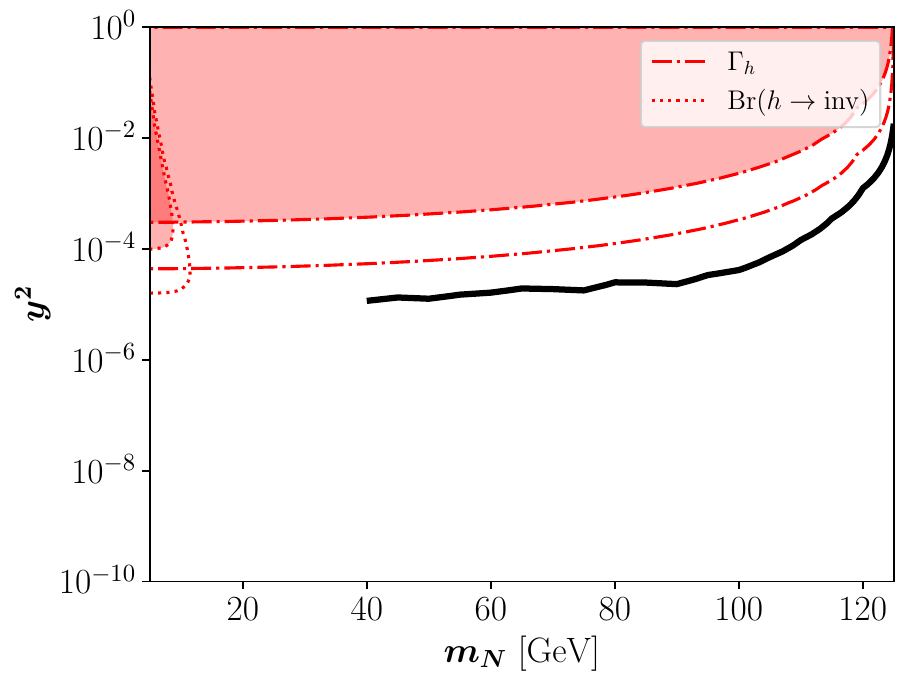}
    \caption{Sensitivity reach of searches for a prompt decay, for $\sqrt{s} = 14$~TeV and $\mathcal{L} = 3$~ab$^{-1}$. The red areas are in tension with the invisible decay of the Higgs or its total decay width, whereas the red dotted and dashed dotted areas correspond to the projected sensitivity at HL-LHC.}
    \label{fig:prompt}
\end{figure}

The discovery significance is given by 
\begin{equation}
   \sigma_3 = \frac{\mathcal{N}_S}{\sqrt{\mathcal{N}_S+\mathcal{N}_B}}\,,
\end{equation}
where $\sigma_3$ denotes the discovery significance at a luminosity of $\mathcal{L} = 3$~ab$^{-1}$, with $\mathcal{N}_S$ and $\mathcal{N}_B$ denoting the corresponding number of signal and background events which fulfill all previous cuts. Figure~\ref{fig:prompt} shows, with a thick solid black line, the sensitivity reach of searches for a prompt decay for $\sqrt{s} = 14$~TeV with the above luminosity, in the plane $[m_N,\, y^2]$. The parameter space above the line could be probed at HL-LHC. In addition, the red areas are in tension with measurements of the total decay width of the Higgs or its branching ratio into invisible, whereas the dotted and dashed dotted areas correspond to the projected sensitivity at HL-LHC.
%%%%%%%%%%%%%%%%%%%%%%%%%%%%%%%%%%%%%%%%%%%%%%%%%%%
%%%%%%%%%%%%%%%%%%%%%%%%%%%%%%%%%%%%%%%%%% 

%%%%%%%%%%%%%%%%%%%%%%%%%%%%%%%%%%%%%%%%%%%%%%%%%%%%%%%%%%
\section{Displaced Vertices} \label{sec:displaced}
\begin{figure}[t!]
    \def\sepf{0.49}
    \centering
    \includegraphics[width=\sepf\columnwidth]{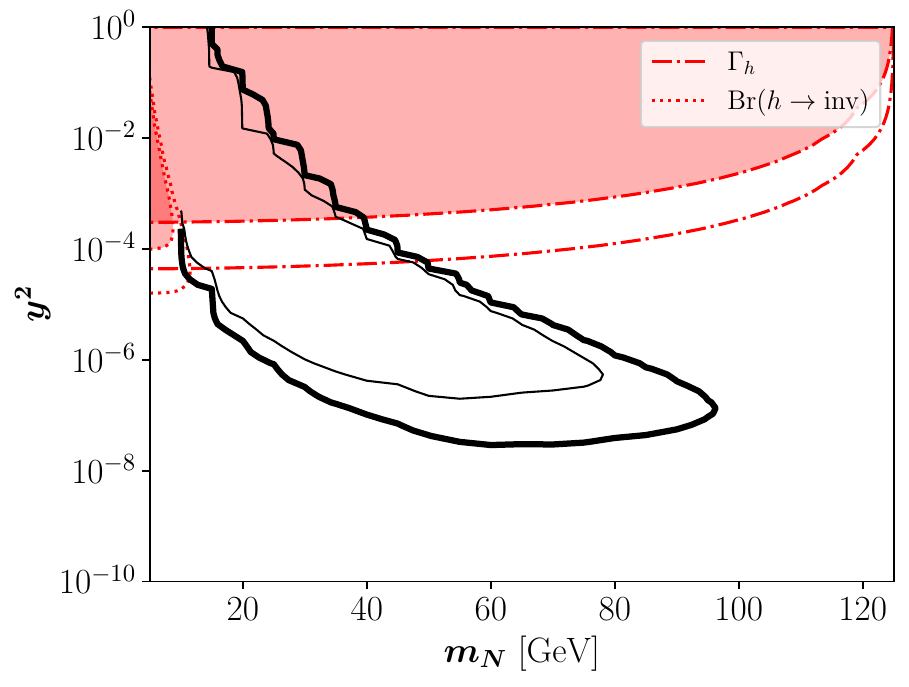}
    \caption{Sensitivity reach of searches for a displaced decay, for $\sqrt{s} = 14$~TeV and $\mathcal{L} = 3$~ab$^{-1}$.
    The thick line corresponds to the assumption of zero background events, whereas for the thin line 65 events were assumed; see the text for further details. The red areas are in tension with the invisible decay of the Higgs or its total decay width, whereas the red dotted and dashed dotted areas correspond to the projected sensitivity at HL-LHC.}
    \label{fig:displaced}
\end{figure}

%%%%%%%%%%%%%%%%%%%%%%%%%%%%%%%%%%%%%%%%%%%%%%%%%%%%%%%%%%
In this section, we focus on the case where HNLs decay in the inner tracker of ATLAS or CMS, featuring a displaced vertex. Here too, we stick to the case where the Higgs is produced through VBF, to trigger on the characteristic two hard forward jets. However, as seen in Section~\ref{sec:Prompt}, ggF events can also have a non-negligible contribution to the same topology requirements and hence have been included in the analysis. For the events, we use the following selection criteria:
\begin{itemize}
    \item Same cuts $S_1$, $S_3$ and $S_4$ for the VBF hard forward jets, as described in Section~\ref{sec:Prompt}.
    \item Two extra jets with a displacement of 1~mm $\leq d_{xy} \leq 1$~m and $d_z \leq 300$~mm.
\end{itemize}
In the first attempt, we assume that there is no background. Therefore, we follow a Poisson distribution and highlight the parameter space where there are more than 3.09 expected signal events at 95\%~CL~\cite{ParticleDataGroup:2022pth}.
Figure~\ref{fig:displaced} shows, with a thick solid black line, the sensitivity reach of searches for a displaced decay for $\sqrt{s} = 14$~TeV and $\mathcal{L} = 3$~ab$^{-1}$, in the plane $[m_N,\, y^2]$.
The rectangular shape of the sensitivity reach can be understood as follows. High values for $m_N$ and $y^2$ cause prompt decays; in contrast, low values generate decays outside the detector. Finally, very small values of the Yukawa coupling cannot be explored due to lack of statistics; see Eq.~\eqref{eq:Ndec}.
In the figure, we also overlay in red the areas that are in tension with current measurements of the total decay width of the Higgs, or its branching ratio into invisible, whereas the red dotted and dashed dotted areas correspond to the projected sensitivity at HL-LHC. Yukawa couplings as small as $y^2 \simeq \mathcal{O}(10^{-8})$ could be probed for the mass range 60~GeV $\lesssim m_N \lesssim 100$~GeV.
%%%%%%%%%%%%%%%%%%%%%%%%%%%%%%%%%%%%%%%%%%%%%%%%%%%
%%%%%%%%%%%%%%%%%%%%%%%%%%%%%%%%%%%%%%%%%% 

The previously used background-free hypothesis could be too optimistic, since background events could come from misidentified displaced events due to detector resolution effects or from random track crossings. To have a rough estimate of how much the inclusion of a background could worsen the sensitivity, we follow the discussion in Refs.~\cite{Deppisch:2018eth, Abada:2018sfh, Marcano:2018fto, Abada:2018ulr}, and assume the largest possible background which is in agreement with not having observed any background event in the ATLAS analysis of Refs.~\cite{ATLAS:2022zhj, ATLAS:2023oti}. This means up to 3 background events for $\mathcal{L} = 139$~fb$^{-1}$, which can be scaled to 65 background events for a luminosity of 3~ab$^{-1}$. This pessimistic scenario is shown in Fig.~\ref{fig:displaced} with a thin black line. This would imply that the HL-LHC could still provide new information on HNLs that display a displaced vertex; in particular, Yukawa couplings of the order $y^2 \simeq \mathcal{O}(10^{-7})$ could be probed if 60~GeV $\lesssim m_N \lesssim 80$~GeV. However, again we stress that a complete background analysis is needed to have more realistic LHC sensitivities.

%%%%%%%%%%%%%%%%%%%%%%%%%%%%%%%%%%%%%%%%%%%%%%%%%%%%%%%%%%
\section{Long-lived Particles} \label{sec:LLP}
%%%%%%%%%%%%%%%%%%%%%%%%%%%%%%%%%%%%%%%%%%%%%%%%%%%%%%%%%%
Long-lived HNLs could be looked for in already approved detectors at the LHC such as FASER~\cite{Feng:2017uoz} and MoEDAL-MAPP1~\cite{Pinfold:2019nqj}. They also have follow-up programs at the high-luminosity (HL) LHC: FASER-2~\cite{FASER:2018eoc} and MoEDAL-MAPP2~\cite{Pinfold:2019zwp}. Other experimental proposals include MATHUSLA~\cite{Chou:2016lxi, Curtin:2018mvb, MATHUSLA:2020uve}, ANUBIS~\cite{Bauer:2019vqk}, CODEX-b~\cite{Gligorov:2017nwh}, and more recently, FA\-CET~\cite{Cerci:2021nlb}. All of these detectors will be sensitive to particles that decay $\mathcal{O}(10)$~m to $\mathcal{O}(500)$~m away from the interaction point (IP). The large distance to the IP guarantees a very low background environment, typically assumed to be negligible. Therefore, only 3.09 events are required to define the sensitivity of the experiment at 95\%~CL~\cite{ParticleDataGroup:2022pth}.

In this section, we will consider HNLs produced by Higgs decays. The total single-Higgs boson inclusive cross section is dominated by the ggF and VBF processes, which is $\sigma \simeq 59.1$~pb at $\sqrt{s} = 14$~TeV~\cite{ggF, VBF}. We will again use a total integrated luminosity $\mathcal{L} = 3$~ab$^{-1}$, and focus on two detectors: FASER and MATHUSLA.

The FASER detector is located at $L = 480$~m downstream of the proton-proton IP used by the ATLAS experiment~\cite{Feng:2017uoz, FASER:2018eoc}. It is sensitive to new particles that decay in a cylindrical volume with radius $R = 10$~cm and length $\Delta = 1.5$~m. In its first phase, it will operate with integrated luminosity $\mathcal{L} = 150~\text{fb}^{-1}$. However, a second phase (FASER-2) is expected to operate with $\Delta = 10$~m, $R = 1$~m and $\mathcal{L} = 3~\text{ab}^{-1}$, at the HL-LHC. Here, we will focus on this second phase.

Given the geometry of FASER, the probability $\mathcal{P}$ of a HNL decaying inside the detector is\footnote{We note that a code for estimating the sensitivities of long-lived particles in different detectors has recently been released~\cite{Domingo:2023dew}; here, however, we use our own implementation.}
\begin{equation}
    \mathcal{P} = \left[e^{-(L - \Delta)/d} - e^{-L/d}\right]\, \Theta\left[R - L\, \tan\theta\right],
\end{equation}
where $d$ is the decay length in the laboratory frame of the HNL, and $\theta$ the angle between the momentum and the beam line. This decay length takes into account the Lorentz boost factor
\begin{equation}
    d = c \tau\, \beta\, \gamma = c \tau\, \frac{|\vec{p}|}{m_N}\,.
\end{equation}

\begin{figure}[t!]
    \def\sepf{0.49}
    \centering
    \includegraphics[width=\sepf\columnwidth]{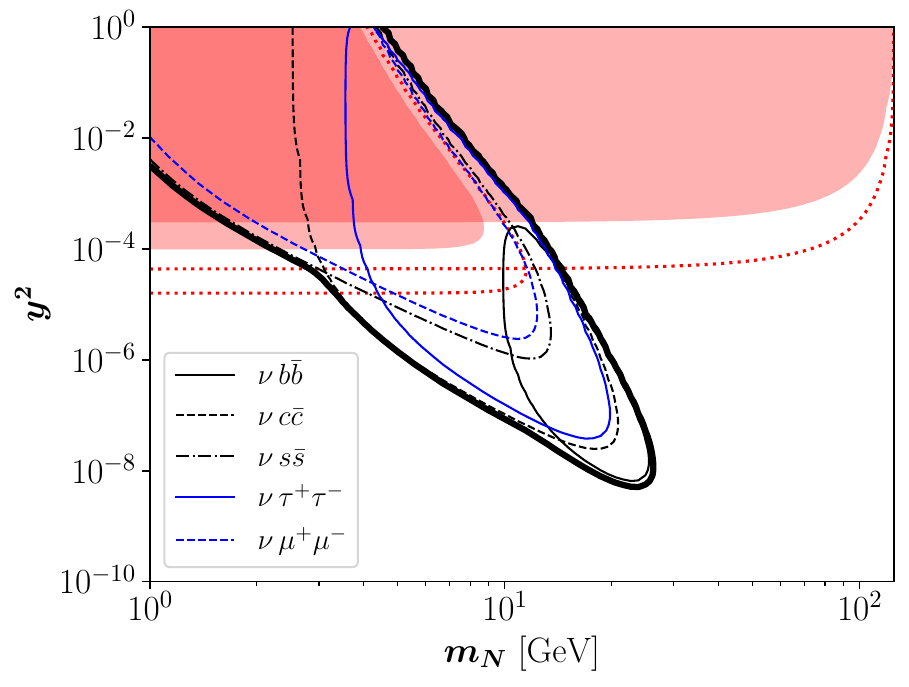}
    \includegraphics[width=\sepf\columnwidth]{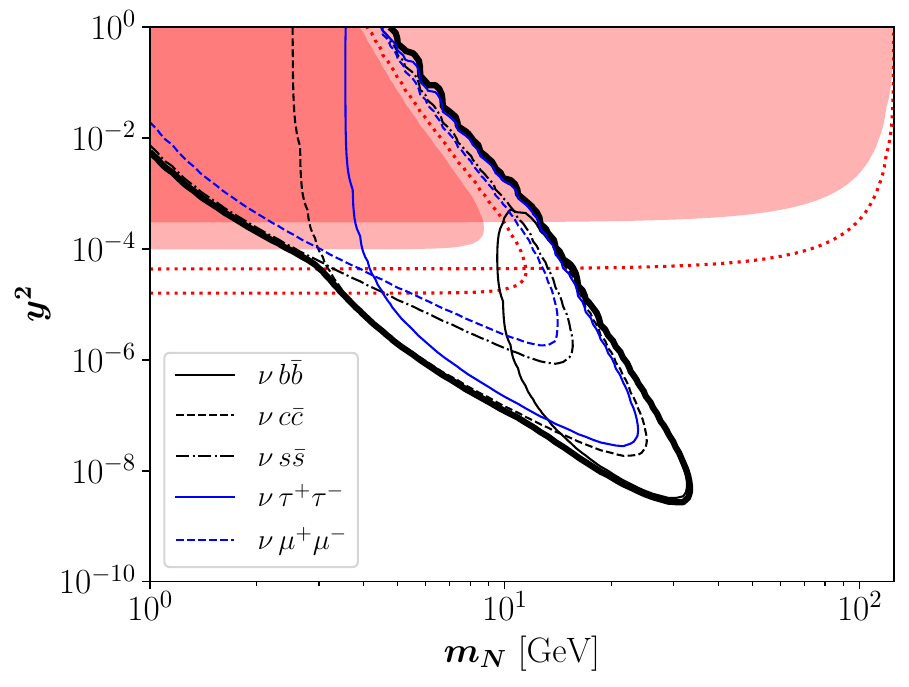}
    \caption{The sensitivity reaches of FASER (left) and MATHUSLA (right) are represented by solid black lines, for $\sqrt{s} = 14$~TeV and $\mathcal{L} = 3$~ab$^{-1}$. The partial contributions of the different channels are shown with thin lines. The red areas are in tension with the invisible decay of the Higgs or its total decay width, whereas the red dotted and dashed dotted areas correspond to the projected sensitivity at HL-LHC.}
    \label{fig:faser}
\end{figure}

The left panel of Fig.~\ref{fig:faser} shows, with a thick solid black line, the sensitivity reach of FASER-2 to HNLs in the plane $[m_N,\, y^2]$. For completeness, the partial contributions of the different channels are also shown with thin lines. As expected, FASER is sensitive to masses smaller than in the cases of prompt decays and displaced vertices because a much longer decay length is required. Additionally, the red areas in the figure are in tension with measurements of the total decay width of the Higgs or its branching ratio into invisible, whereas the red dotted and dashed dotted areas correspond to the projected sensitivity at HL-LHC. It is interesting to note that the experiment will be sensitive to unconstrained regions of the parameter space corresponding to masses between $\sim 3$~GeV and $\sim 30$~GeV and couplings as low as $y^2 \simeq \mathcal{O}(10^{-8})$.
%%%%%%%%%%%%%%%%%%%%%%%%%%%%%%%%%%%%%%%%%%%%%%%%%%%
%%%%%%%%%%%%%%%%%%%%%%%%%%%%%%%%%%%%%%%%%% 

MATHUSLA~\cite{MATHUSLA:2020uve} is a box-shaped 100~m $\times$ 100~m $\times$ 25~m far detector for the CMS interaction point. Taking the CMS IP to be located at $x = y = z = 0$, its front and depth are parallel to the beam
axis, at distances $z_\text{min} = 68$~m and $z_\text{max} = 168$~m.  Its base and cover are located at $y_\text{min} = 60$~m and $y_\text{max} = 85$~m, above the beam. The boundaries on the left and right sides are slightly offset, $x_\text{min} = -42.41$~m and $x_\text{max} = 57.59$~m.
The right panel of Fig.~\ref{fig:faser} compares to the left panel, but for the case of MATHUSLA. We note that the sensitivity reach for both experiments is comparable, although MATHUSLA could probe a slightly larger parameter space and smaller Yukawa couplings $y^2 \simeq \mathcal{O}(10^{-9})$.

%%%%%%%%%%%%%%%%%%%%%%%%%%%%%%%%%%%%%%%%%%%%%%%%%%%%%%%%%%
\section{Conclusions} \label{sec:concl}
%%%%%%%%%%%%%%%%%%%%%%%%%%%%%%%%%%%%%%%%%%%%%%%%%%%%%%%%%%
The introduction of heavy neutral leptons (HNLs) is a natural and minimal extension of the Standard Model (SM) that can play a fundamental role in the solution of long standing problems such as the generation of neutrino masses, the baryon asymmetry of the universe and the dark matter. Many experimental searches have been conducted with the aim of discovering HNLs, mainly, if not exclusively, through its potential mixing with active SM neutrinos.

%%%%%%%%%%%%%%%%%%%%%%%%%%%%%%%%%%%%%%%%%%%%%%%%%%%
\begin{figure}[t!]
    \def\sepf{0.9}
    \centering
    \includegraphics[width=\sepf\columnwidth]{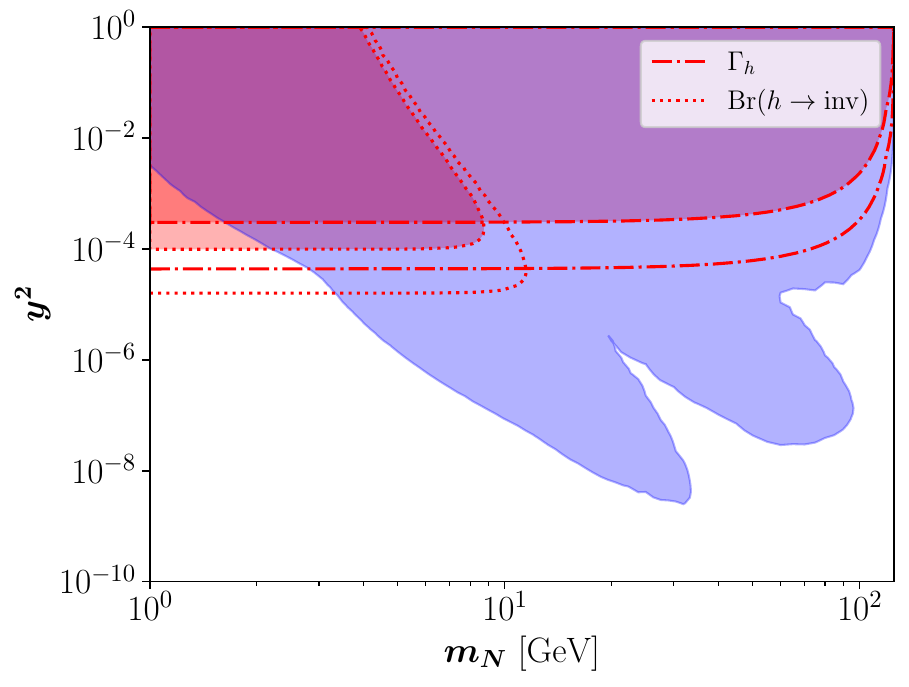}
    \caption{The combined sensitivity reach for searches of prompt decays, displaced vertices in the tracker and decays in far detectors (FASER and MATHUSLA), for $\sqrt{s} = 14$~TeV and $\mathcal{L} = 3$~ab$^{-1}$, is represented in blue. The red areas are in tension with the invisible decay of the Higgs or its total decay width, whereas the red dotted and dashed dotted areas correspond to the projected sensitivity at HL-LHC.}
    \label{fig:all}
\end{figure} 
%%%%%%%%%%%%%%%%%%%%%%%%%%%%%%%%%%%%%%%%%% 
Alternatively, here, we have focused on a novel and complementary approach, exploiting the Yukawa interaction between HNLs and the SM Higgs boson. Even if a relation can be expected between the mixing and the Yukawa couplings, it is model dependent and therefore cannot be taken for granted. With this in mind, in this first study we have ignored potential large mixings and analyzed the reach of HL-LHC to discover HNLs through their Yukawa coupling to the SM Higgs.

In this scenario, HNLs are produced by the decay of Higgs bosons. Depending on its mass and Yukawa coupling, HNLs can decay promptly inside ATLAS or CMS detectors, feature a displaced vertex in the tracker, or even decay far away from the interaction point in far detectors such as FASER or MATHUSLA. The sensitivity reach of each of these options is presented in Figs.~\ref{fig:prompt}, \ref{fig:displaced}, and~\ref{fig:faser}, respectively. To facilitate comparison, Fig.~\ref{fig:all} summarizes in blue the combined sensitivity reach for $\sqrt{s} = 14$~TeV and $\mathcal{L} = 3$~ab$^{-1}$. Note that the red areas are in tension with current measurements of the total decay width of the Higgs boson or its branching ratio into invisible. In Fig.~\ref{fig:all}, we also present the projected sensitivity obtained from the invisible decay of the Higgs boson and the future measurement of the total decay width of the Higgs boson at HL-LHC~\cite{deBlas:2022aow}. HL-LHC combined with future Higgs factories have the potential to reach precision up to $1\%$ for the branching corresponding to invisible decay of the Higgs boson and up to $1.1\%$ for the total width of the Higgs boson~\cite{deBlas:2022aow}, which roughly translates to a value of $y^2 \sim 9\times 10^{-6}$.

%%%%%%%%%%%%%%%%%%%%%%%%%%%%%%%%%%%%%%%%%%%
\acknowledgments
%%%%%%%%%%%%%%%%%%%%%%%%%%%%%%%%%%%%%%%%%%%
We thank Giovanna Cottin, Marco Drewes, Chee Sheng Fong and Avelino Vicente for valuable discussions.

%%%%%%%%%%%%%%%%%%%%%%%%%
\appendix
%%%%%%%%%%%%%%%%%%%%%%%%%
\section{Appendix} \label{appendix}
%%%%%%%%%%%%%%%%%%%%%%%%%%%%%%%%%%%%%%%%%%%%%%
The decoupling between the Yukawa coupling between SM neutrinos and HNL can be achieved if we allow the HNLs to be Dirac particles ($\nu_L$, $\nu_R$) and prevent tree-level Dirac mass terms by introducing additional symmetries. By incorporating a heavy Dirac HNL ($N_L$, $N_R$) and new complex scalar singlets with specific charges under the new symmetries such that only $N_R$ couples to the light neutrino ($\nu_L$) via the SM Higgs and $N_L$ couples to $\nu_R$ via a new complex scalar singlet, we achieve two distinct Dirac masses in the neutrino-mass matrix.  Refs.~\cite{Ma:2014qra, CentellesChulia:2016rms} achieve this by extending the SM gauge group with a $\mathbb{Z}_4 \otimes \mathbb{Z}_2$ and a $U(1)_{B-L}$ respectively. The part of the Lagrangian for one generation of neutrinos that holds relevance in the current context can be written as:
\begin{equation}
    \mathcal{L}\supset y \Bar{L}_L \Tilde{H} N_R + g \Bar{N}_L \chi \nu_R + M \Bar{N}_L N_R
\end{equation}
with $\langle H \rangle = (0,\, v_h/\sqrt{2})^T$, $\langle\chi\rangle = v_\chi$ and $\tilde H = i\, \tau_2\, H^*$, with $\tau_2$ the second Pauli matrix and $v_h \simeq 246$~GeV. After the Higgs ($H$) and the new scalar ($\chi$) get a vev, the ensuing  mass matrix can be written as
\begin{equation}
   \mathcal{M}=
    \begin{pmatrix}
        0 & m_1 \\
        m_2 & M 
    \end{pmatrix}    
\end{equation}
where $m_1 \equiv y\, v_h/\sqrt{2}$ and $m_2 \equiv g\, v_\chi$.
The mass matrix can be diagonalized through $\mathcal{M}_D = P^{-1}\, \mathcal{M}\, P$, where
\begin{equation}
    P =
    \begin{pmatrix}
        \frac{2 m_1}{\sqrt{4\, m_1^2 + \left(M - \sqrt{M^2 + 4\, m_1\, m_2}\right)^2}} & \frac{2 m_1}{\sqrt{4\, m_1^2 + \left(M + \sqrt{M^2 + 4\, m_1\, m_2}\right)^2}} \\
        \frac{\left(M - \sqrt{M^2 + 4\, m_1\, m_2}\right)}{\sqrt{4\, m_1^2 + \left(M - \sqrt{M^2 + 4\, m_1\, m_2}\right)^2}} & \frac{\left(M + \sqrt{M^2 + 4\, m_1\, m_2}\right)}{\sqrt{4\, m_1^2 + \left(M +\sqrt{M^2 + 4\, m_1\, m_2}\right)^2}}
    \end{pmatrix}    
\end{equation}
and the diagonal matrix with the two eigenvalues denoted by
\begin{equation}
    \mathcal{M}_D = 
    \begin{pmatrix}
        \frac12 \left[M - \sqrt{M^2 + 4\, m_1\, m_2}\right] & 0 \\
        0 & \frac12 \left[M + \sqrt{M^2 + 4\, m_1\, m_2}\right] 
    \end{pmatrix}.
\end{equation}
The flavor states and mass states are related through
\begin{equation}
    \begin{pmatrix}
       \Bar{\nu}_L \\ \Bar{N}_L
    \end{pmatrix} = (P^{-1})^{T} 
    \begin{pmatrix}
       \Bar{\nu}^m_L \\ \Bar{N}^m_L
    \end{pmatrix}, ~ \begin{pmatrix}
       {\nu}_R \\ {N}_R
    \end{pmatrix} = P
    \begin{pmatrix}
       {\nu}^m_R \\ {N}^m_R
    \end{pmatrix};
\end{equation}
where $(\Bar{\nu}^m_L ~~ \Bar{N}^m_L)^T $ and $({\nu}^m_R ~~ {N}^m_R)^T $ are the mass eigenstates. In the limit of $m_2$ becoming negligible, these relations become
\begin{equation}
   \begin{pmatrix}
       \Bar{\nu}_L \\ \Bar{N}_L
    \end{pmatrix} = 
    \begin{pmatrix}
       \Bar{\nu}^m_L \\ \frac{\sqrt{m_1^2 + M^2}}{M}\Bar{N}^m_L - \frac{m_1}{M} \Bar{\nu}^m_L
    \end{pmatrix}, ~~
     \begin{pmatrix}
       {\nu}_R \\ {N}_R
    \end{pmatrix} = 
    \begin{pmatrix}
       {\nu}^m_R + \frac{m_1}{\sqrt{m_1^2 + M^2}}{N}^m_R \\ \frac{M}{\sqrt{m_1^2 + M^2}}{N}^m_R 
    \end{pmatrix}.
\end{equation}
In this limit, $\nu_L$ and $N_R$, which provide the Yukawa term with the Higgs, are pure states and hence do not have any mixing term. However, mixings appear for nonvanishing $m_2$. In the limit $m_2 \ll M$, one has
\begin{equation}
   \begin{pmatrix}
       \Bar{\nu}_L \\
       \Bar{N}_L
    \end{pmatrix} = 
    \begin{pmatrix}
       \left[1 - \frac{m_1\, m_2}{M^2}\right] \Bar{\nu}^m_L + \frac{m_2 \sqrt{m_1^2 + M^2}}{M^2} \Bar{N}^m_L\\
        - \frac{m_1}{M} \Bar{\nu}^m_L + \frac{\sqrt{m_1^2 + M^2}}{M}\Bar{N}^m_L
    \end{pmatrix}, ~~
     \begin{pmatrix}
       {\nu}_R \\ {N}_R
    \end{pmatrix} = 
    \begin{pmatrix}
       \left[1 - \frac{m_2^2}{2 M^2}\right]{\nu}^m_R + \frac{m_1}{\sqrt{m_1^2 + M^2}}{N}^m_R \\
       - \frac{m_2}{M}\, {\nu}^m_R + \frac{M}{\sqrt{m_1^2 + M^2}}{N}^m_R 
    \end{pmatrix}.
\end{equation}
The charged and neutral current interactions of the SM gauge bosons with the light neutrinos ($\nu_L$) become:
\begin{align}
    \mathcal{L}_{cc}^{\nu_L} &\sim -\frac{g}{2\sqrt{2}}\Bigg\{ \left(1 - \frac{m_1\, m_2}{M^2}\right)\Big[(\Bar{\nu}^m_L \gamma^\mu l_L)~W_\mu^\dagger \Big]+\frac{m_2 \sqrt{m_1^2 + M^2}}{M^2} \Big[(\Bar{N}^m_L \gamma^\mu l_L)~W_\mu^\dagger\Big]+ H.c.\Bigg\}, \label{new-cc} \\
    \mathcal{L}_{nc}^{\nu_L} &\sim -\frac{e}{ 2 \sin{\theta_W} \cos{\theta_W}} \Bigg\{\left(1 - \frac{2~m_1\, m_2}{M^2}\right) \Big[(\Bar{\nu}^m_L \gamma^\mu \nu_L^m)~Z_\mu\Big] \nonumber \\
    &\qquad + \frac{2 m_2 \sqrt{m_1^2 + M^2}}{M^2} \Big[(\Bar{\nu}^m_L \gamma^\mu N_L^m)~Z_\mu\Big] + \frac{m_2^2 (m_1^2 + M^2)}{M^4} \Big[(\Bar{N}^m_L \gamma^\mu N_L^m)~Z_\mu\Big] \Bigg\}. \label{new-nc}
\end{align}
Similarly, the Yukawa term with the physical SM Higgs boson ($h$) can now be written as:
\begin{equation}\label{new-yuk}
\begin{split}
    \mathcal{L}_{Yuk}^{\nu_L} \sim \frac{y}{\sqrt{2}} \Bigg\{\left(1 - \frac{m_1\, m_2}{M^2}\right) \left(\frac{M}{\sqrt{m_1^2 + M^2}}\right) \Big[\Bar{\nu}^m_L  N_R^m~h\Big] - \left(\frac{m_2}{M}\right)\Big[\Bar{\nu}^m_L  \nu_R^m~h\Big] \\
    -\left(\frac{m_2^2 \sqrt{m_1^2 + M^2}}{M^3}\right)\Big[\Bar{N}^m_L  \nu_R^m~h\Big]
    + \left(\frac{m_2}{M}\right)\Big[\Bar{N}^m_L  N_R^m~h\Big]\Bigg\}. 
\end{split}    
\end{equation}
It is clear from Eqs.~\eqref{new-cc}, \eqref{new-nc}, and~\eqref{new-yuk} that the production and decay of $N_R^m$ is only possible through the Yukawa interaction with $h$ and $\nu_L^m$ at tree level, while the SM gauge-boson interactions through mixing do not play a role in it (or in other words, the mixing is precisely zero). For small $m_2$ values, only the first term of Eq.~\eqref{new-yuk} will contribute to the phenomenology related to $N_R$. In contrast, $N_L$ couples to the gauge bosons through mixing and can have potentially higher production and decay rates compared to the Yukawa interactions.

%%%%%%%%%%%%%%%%%%%%%%%%%
\bibliographystyle{JHEP}
\bibliography{biblio}
%%%%%%%%%%%%%%%%%%%%%%%%%
\end{document}